\documentclass[prl,showpacs,floatfix,twocolumn]{revtex4}
\usepackage{graphicx}
\usepackage{dcolumn}

\begin{document}

\newcommand*{\cm}{cm$^{-1}$\,}
\newcommand*{\nco}{Na$_x$CoO$_2$\,}
\newcommand*{\eg}{e$_g$\,}
\newcommand*{\tg}{t$_{2g}$\,}

%
\title{Infrared probe of the electronic structure and charge dynamics in Na$_{0.7}$CoO$_2$}
%
%

\author{N. L. Wang}
\email{nlwang@aphy.iphy.ac.cn}%
\affiliation{Institute of Physics and Center for Condensed Matter
Physics, Chinese Academy of Sciences, P.~O.~Box 603, Beijing
100080, P.~R.~China}
\author{P. Zheng}
\author{D. Wu}
\author{Y. C. Ma}
\affiliation{Institute of Physics and Center for Condensed Matter
Physics, Chinese Academy of Sciences, P.~O.~Box 603, Beijing
100080, P.~R.~China}
\author{T. Xiang}
\affiliation{Institute of Theoretical Physics and
Interdisciplinary Center of Theoretical Studies, Chinese Academy
of Sciences, Beijing 100080, P.~R.~China}
\author{R. Y. Jin }
\author{D. Mandrus}

\affiliation{Solid State Division, Oak Ridge National Laboratory,
Oak Ridge, TN 37831}%
%
%
%
\begin{abstract}
We present measurements of the optical spectra on
Na$_{0.7}$CoO$_2$ single crystals. The optical conductivity shows
two broad interband transition peaks at 1.6 eV and 3.1 eV, and a
weak midinfrared peak at 0.4 eV. The intraband response of
conducting carriers is different from that of a simple Drude
metal. A peak at low but finite frequency is observed, which
shifts to higher frequencies with increasing temperature, even
though the dc resistivity is metallic. The origin of the interband
transitions and the low-frequency charge dynamics have been
discussed and compared with other experiments.

\end{abstract}


\maketitle

%

The recent discovery of superconductivity at $\sim$5 K in hydrated
sodium cobaltate \cite{Takada} has attracted much attention, as it
is a new system other than cuprates where a doped Mott insulator
becomes a superconductor. The host compound \nco consists of
alternate stacking of Na and CoO$_2$ layers in which edge sharing
CoO$_6$ octahedra lead to a two-dimensional (2D) triangular
lattice of Co ions.\cite{Terasaki1,Terasaki2} Superconductivity
occurs when Na content x is near 0.3 and sufficient water is
intercalated between the CoO$_2$ layers. This material provides a
model system for studying the physics of correlated electrons in a
2D triangular lattice. It is also believed that the study of \nco
system may shed new light on high-temperature superconductivity in
cuprates.

To understand the mechanism of superconductivity in this material,
great effort has been paid to the investigation of its electronic
structure. As the first step, attention has been given to the host
compound. A detailed band structure calculation for x=0.5 compound
has been performed by Singh.\cite{Singh1} Within the local density
approximation (LDA), it was found that the O 2p bands are located
well below the Fermi level and there is little hybridization
between O 2p and Co 3d bands. The splitting between the Co 3d \eg
and \tg bands is very large. The conduction electrons are mainly
from the \tg band and the \eg band is about 2.5 eV above the Fermi
level. In the rhombohedral crystal field, the \tg manifold is
further split into two a$_g$ and four e$_g^\prime$ bands. The
a$_g$ bands contribute a large cylindrical hole Fermi surface
centered at $\Gamma$. This large Fermi surface was confirmed by
the ARPES measurements \cite{Valla,Hasan,Yang}. However, the six
Fermi pockets from the e$_g^\prime$ bands, predicted by the LDA,
were not observed by the ARPES. On the other hand, if the effects
of spin polarization were included, an itinerant ferromagnetic
state was predicted to exist for x=0.3 to
0.7.\cite{Singh1,Singh2,Kunes} There was no report on the
existence of distinct magnetic order in this material, but a
Curie-Weiss-like susceptibility $\chi(T)$ was found when x
$\approx$ 0.7. This suggests that Co ions have local moments
associated with the exchange
splitting\cite{Takada,Sakurai,Lorenz}, but strong quantum
fluctuations suppress the long-range ferromagnetic ordering in
\nco.\cite{Singh2}

In this work, we report the in-plane optical response over broad
frequencies on single crystals with x=0.7. Our measurements yield
two broad interband transition peaks at 1.6 eV and 3.1 eV. The
interband transition energies are significantly different from the
energy difference between the occupied \tg and empty \eg bands
($\sim$2.5 eV) as predicted by the LDA. The physical origin for
this difference will be discussed. Moreover, our experiments also
revealed a weak midinfrared peak at 0.4 eV and unusual charge
dynamics at low frequencies.

High-quality Na$_{0.7}$CoO$_2$ single crystals with size around
3mm$\times$3mm were grown using the flux method.\cite{Jin} Fig. 1
shows the in-plane dc resistivity $\rho_{ab}$ measured by
four-contact method. The temperature dependence of $\rho_{ab}$ is
typically metallic and agrees with those reported in Ref.
\cite{Hasan,Wang,Foo} with similar Na concentration.

\begin{figure}[t]
\centerline{\includegraphics[width=2.6in]{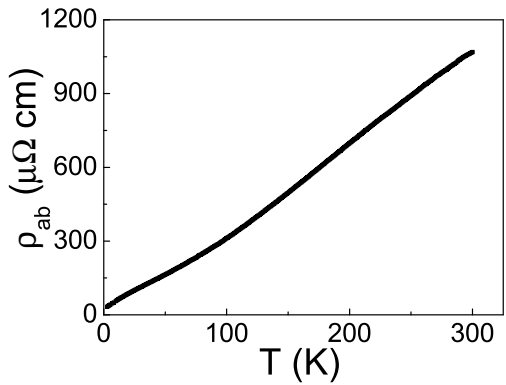}}%
\vspace*{0cm}%
\caption{The in-plane dc resistivity of Na$_{0.7}$CoO$_2$ single crystal as a function of temperature.}%
\label{fig1}
\end{figure}

The crystals could be readily cleaved to obtain a fresh and shinny
surface. The near-normal incident reflectance spectra were
measured by a Bruker 66v/S spectrometer in the frequency range
from 40 \cm to 29,000 \cm. The sample was mounted on an optically
black cone in a cold-finger flow cryostat. An \textit{in situ}
overcoating technique was employed for reflectance measurements
\cite{Homes}. The optical conductivity spectra were obtained from
a Kramers-Kronig transformation of R($\omega$). We use
Hagen-Rubens' relation for the low frequency extrapolation, and a
constant extrapolation to 100,000 \cm followed by a well-known
function of $\omega^{-4}$ in the higher-energy side.

Fig. 2 shows the in-plane reflectance spectra measured at
different temperatures between 10 K and 300 K. In accord with the
metallic dc resistivity behavior with a positive slope, we found
that the low-frequency reflectance increases with decreasing
temperature. However, the reflectance in the mid-infrared region
decreases with decreasing temperature and $R(\omega)$ at different
temperatures cross at about 1,300 \cm. The main spectral change
caused by temperature variation is roughly below 4,000 \cm. The
reflectance reaches a minimum (or an edge ) near 6,000 \cm. This
frequency is close to the so-called screened plasma frequency. We
note that this frequency is significantly lower than that of
optimally doped cuprates like YBa$_2$Cu$_3$O$_{7-\delta}$ and
Bi$_2$Sr$_2$CaCu$_2$O$_{8+\delta}$, suggesting that the conducting
carrier density is lower in the present compound.

Fig. 3 shows the conductivity spectra over broad frequencies.
Corresponding to the reflectance minimum, the optical conductivity
also exhibits a minimum at similar frequency. The low-frequency
part is mainly contributed by the conducting carriers. A sum of
the spectral weight up to 6,000 \cm yields approximately the
overall plasma frequency of $\sim$ 1.2$\times$10$^4$ \cm, which is
apparently lower than the values for optimally doped cuprates
obtained by the same method.\cite{Puchkov} At high frequency side,
two prominent interband transition peaks centered at 25,000 \cm
(3.1 eV) and 13,000 \cm (1.6 eV) are observed and labelled as
$\alpha$ and $\beta$. A week shoulder is seen for the $\beta$ peak
near 9,000 \cm. The peak strengths, especially the $\alpha$ one,
could be affected by the high-frequency extrapolation in the
Kramers-Kronig transformation, but their shapes and positions are
less affected. In fact, as shown in the inset of Fig. 2, two
interband transition peaks in the measured reflectance in the
above mentioned energy scales can be clearly seen.

\begin{figure}[t]
\centerline{\includegraphics[width=3.1in]{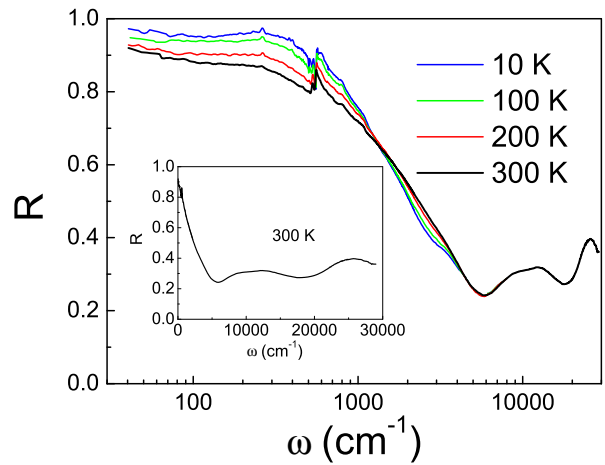}}%
\vspace*{0cm}%
\caption{(Color online) Frequency dependences of the in-plane
reflectance spectra at different temperatures.
The inset is the room temperature reflectance in a linear frequency scale.}%
\label{fig2}
\end{figure}

According to the band structure calculation,\cite{Singh1} the O 2p
bands are much below the Fermi level, and the transition between
the O 2p bands and Co 3d bands is beyond our measurement
frequencies. A most probable interband transition relevant to our
observation is from the occupied Co3d \tg to the empty Co3d \eg
bands due to the octahedral crystal field splitting. However, the
LDA result for the energy difference between the \tg and \eg bands
is 2.5 eV,\cite{Singh1} which lies between those two peaks. We
suggest that the difference could be due to the neglect of the
spin polarization effect in the simple LDA calculation. Within
spin-polarized local density approximation (LSDA),
Na$_{0.5}$CoO$_2$ is expected to be a ferromagnetic half
metal,\cite{Singh1} and the ferromagnetic instability is robust
with respect to doping.\cite{Singh2,Kunes} It is predicted that
the exchange splitting of the \tg states is about 1.5 eV for
Na$_{0.3}$CoO$_2$.\cite{Kunes} Experimentally, although no long
range ferromagnetic ordering was observed, a Curie-Weiss-like
temperature dependence of susceptibility $\chi(T)$ was found for x
near 0.7.\cite{Takada,Sakurai,Lorenz} Recent neutron inelastic
scattering also indicated ferromagnetic spin fluctuations within
the cobalt-oxygen layers for Na$_{0.75}$CoO$_2$.\cite{Boothroyd}
Therefore, the exchange splittings for the spin-up and spin-down
energy levels of \tg and \eg are still expected to exist. This
splitting will broaden the \tg and \eg band widths. On one hand,
it reduces the minimum excitation gap between the \tg and \eg
bands (leading to the $\beta$ peak); on the other hand, it leads
to further increase of the largest energy separation between the
\tg and \eg bands ($\alpha$ transition). So, the two peaks
resemble the respective transitions between spin minority and spin
majority DOS as observed in ferromagnetic materials, such as
CrO$_2$.\cite{Singley,Mazin}

\begin{figure}[t]
\centerline{\includegraphics[width=3.1in]{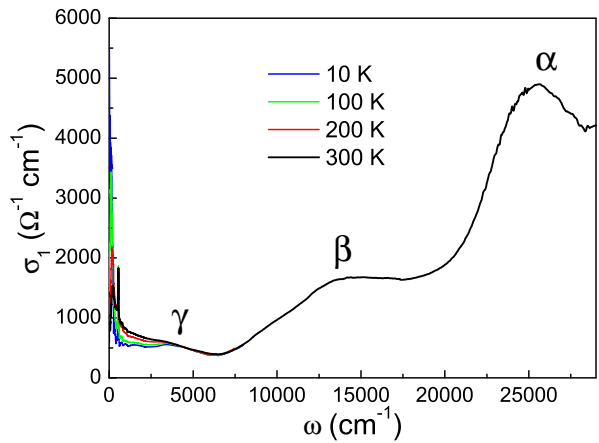}}%
\vspace*{0.0cm}%
\caption{(Color online) The in-plane optical conductivity spectra
of Na$_{0.7}$CoO$_2$ over broad range of frequencies. $\alpha$ and
$\beta$ represent two interband transitions, $\gamma$ is a weak absorption feature in the mid-infrared region.}%
\label{fig3}
\end{figure}

In addition to the interband transition, there is a weak
absorption feature (labelled as $\gamma$ in fig. 3) in the
mid-infrared region around $\sim$ 3,300 \cm (0.4 eV). This feature
is already present at room temperature, but becomes more
pronounced at low temperatures. There are several possibilities
for the origin of this absorption feature. A simple explanation is
that the weak peak is due to the interband transition between
different \tg states. The ARPES experiments along the $\Gamma$-K
direction revealed the existence of a nearly flat band at 0.16 -
0.2 eV below Fermi energy and an extended flat band just above
E$_F$. \cite{Yang} However, the transition between them is
unlikely to be responsible for this feature, since the excitation
energy is smaller than 0.4 eV. Thus it is more likely that this
weak peak is dut to a transition from occupied e$_g^\prime$ bands
to some partially filled a$_g$ bands within \tg manifold. An
alternative explanation is that the weak mid-infrared feature is
due to strong correlation effects. Recent theoretical study based
on a fermion-spin theory of the t-J model on a triangular
lattice\cite{Liu} shows that, besides a Drude-like peak at lower
frequencies, a mid-infrared peak exists in this material. In this
picture, the mid-infrared peak results from the competition
between the kinetic and magnetic exchange energies. It is also
possible that this peak results from spin polarons induced by
magnetic fluctuations.

Fig. 4 shows the conductivity spectra below 1,500 \cm. The low
frequency conductivity increases with decreasing temperature,
consistent with the metallic behavior of the dc resistivity shown
in Fig. 1. The increase of the spectral weight at low frequencies
comes from the region below the mid-infrared peak. At room
temperature, two in-plane phonons could be seen clearly at 525 and
553 \cm. With decreasing temperature, they shift slightly towards
higher frequencies, and two additional phonons become visible at
505 and 575 \cm. In Na$_x$CoO$_2$, Na ions have two possible
sites, leading to two kinds of structural
geometries.\cite{Jorgensen} Symmetry analysis for each geometry
reveals two infrared-active in-plane phonon modes
2E$_{1u}$.\cite{Li} The respective mode frequencies are somewhat
different for the two geometries. The observation of two in-plane
phonon modes at room temperature, which are all close to the
frequency of hard E$_{1u}$ mode, may indicate that the two
structural geometries, or both Na sites occupations, are present
in the sample. The appearance of additional phonon modes might be
due to a structural change with temperature. However, a complete
understanding of the infrared active phonons requires further
efforts.

\begin{figure}[t]
\centerline{\includegraphics[width=3.2in]{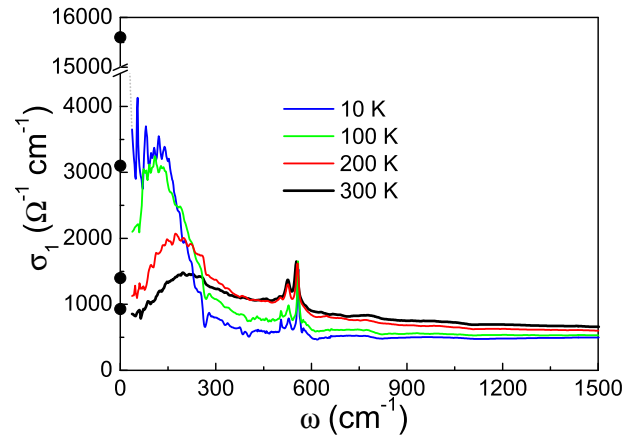}}%
\vspace*{0cm}%
\caption{(Color online) The low-$\omega$ optical conductivity
spectra of Na$_{0.7}$CoO$_2$ at different temperatures. The filled
circles represent the DC conductivity values at 300 K, 200 K, 100
K and 10 K (from bottom to top). The sharp increase
indicated by the grey dash-dot line at 10 K is beyond our lowest measurement frequency.}%
\label{fig4}
\end{figure}

The most striking feature is that the intraband response of
conducting carriers is rather different from the simple Drude
behavior. The conductivity drops at very low frequencies,
resulting in a low frequency peak. The peak shifts toward lower
frequency with decreasing temperature. Such peak has been observed
by Bernhard et al. on Na$_{0.82}$CoO$_2$ single crystals in a very
recent report.\cite{Bernhard} Similar finite energy peaks that
shift to lower energies with decreasing T are predicted for Raman
scattering in correlated systems near metal-insulator transition
on the metallic side.\cite{Freericks} Usually, the drop of optical
conductivity at low frequency is considered as a signature of
charge localization. For \nco, however, this drop is unlikely to
be due to the weak localization effect, since the dc resistivity
is purely metallic and does not show any sign of localization in
the whole temperature range measured.

More information about the low-frequency charge dynamics could be
obtained by comparing the dc resistivity with infrared
conductivity at zero frequency limit. We found that, at room
temperature, the optical conductivity in the low frequency limit
agrees well with the dc data within experimental errors. However,
with decreasing temperature, the dc conductivity data appear to
have higher values. In particular, at 10 K, the dc conductivity is
as high as 15,600 $(\Omega cm)^{-1}$, four times higher than the
low-frequency optical conductivity data. This implies that there
should exist a sharp increase in the optical conductivity below
the frequency limit of our measurements. In this case, the
conductivity drop tends to disappear due to the emergence of the
new component at extremely low frequencies.

We believe that the temperature dependent behavior of the
low-lying excitations observed in optical conductivity has close
relationship with the quasiparticle dynamics observed in ARPES
experiments. For Na$_{0.7}$CoO$_2$ single crystals, a well defined
quasiparticle was observed only at low temperature where the
resistivity is linear in T.\cite{Hasan} The quasiparticle weight
decreases to almost zero on raising temperature to above 100 K. In
a study on Na$_{0.5}$CoO$_2$ crystals, Valla et al.\cite{Valla}
also found that the in-plane quasiparticles exist only at low
temperature where the c-axis transport becomes metallic. If we
compare the infrared data with ARPES results, it becomes clear
that the low-$\omega$ drop in optical conductivity of \nco
correlates with the incoherent electronic states in which the
quasiparticle picture breaks down. It also turns out that the
positive slope of dc transport does not necessarily mean the
existence of well-defined quasiparticles. The high temperature
transport in \nco is incoherent in nature. It is worth pointing
out that the finite-energy peak has also been observed in other
strongly correlated systems.\cite{Takenaka} The coexistence of the
"metallic" dc resistivity with a finite-energy peak in optical
conductivity and the absence of well-defined quasiparticles in
ARPES is a great challenge to our understanding of the charge
transport.

Our results also imply that the development of an extremely narrow
and sharp Drude component at low temperature might correspond to
the well-defined quasiparticles in ARPES. Note that the occurrence
of a sharp and narrow resonance mode at $\omega$=0 in a broad
spectral background at low temperature was widely observed in
strongly correlated electron systems, which was explained in terms
of the renormalizaton of both the effective mass and scattering
time of quasiparticles in the many-body picture.\cite{Degiorgi} On
this basis, this renormalized quasiparticle peak should be
responsible for the moderate mass enhancement observed in the
specific heat measurement for the material.\cite{Terasaki2} The
very narrow Drude peak was also inferred in the very recent work
by Bernhard et al.\cite{Bernhard} from the ellipsometry
measurement on Na$_{0.82}$CoO$_2$. From the decrease of the real
part of dielectric function at very low frequency, they estimated
the plasma frequency for this narrow Drude carriers of roughly
1,300 \cm. This value is almost 10 times smaller than that the
overall plasma frequency obtained from the sum rule in the present
work. Such difference also reflects the enhanced effective mass of
quasiparticle at low temperature.\cite{Degiorgi}

To conclude, the electronic structure and charge dynamics have
been investigated for Na$_{0.7}$CoO$_2$ by means of optical
spectroscopy probe. The optical conductivity spectra show two
broad interband transition peaks at 1.6 eV and 3.1 eV, and a weak
midinfrared peak at 0.4 eV. The former two peaks were explained as
transitions between occupied \tg to empty \eg bands by invoking
the effect of exchange splitting. The midinfrared peak is
attributed either to the interband transition within the \tg
manifold or to the electronic correlation effect. The intraband
response of conducting carriers is different from that of a simple
Drude metal. The optical conductivity exhibits a peak at finite
frequency, which shifts slightly towards higher frequencies with
increasing temperature. Our analysis suggests that the peak is not
due to the carrier localization but represents the incoherent
electronic states where the quasiparticle picture breaks down.

We acknowledge helpful discussions with Z. Fang, S. P. Feng, J. L.
Luo, Y. Wang, L. Yu, and G. M. Zhang. This work is supported by
National Science Foundation of China (No. 10025418, 10374109), the
Knowledge Innovation Project of Chinese Academy of Sciences. Oak
Ridge National Laboratory is managed by UT-Battelle, LLC, for the
U.S. Department of Energy under contract DE-AC05-00OR22725.

%
%

\end{document}